\documentclass[10pt,twoside]{article}

\usepackage{graphicx}
\usepackage{psfig}
\makeindex

\begin{document}

\title{Oscillating neutron star as a pulsar}

\markboth{A.N.\,Timokhin}{Oscillating neutron star as a pulsar}

\author{A.N.\,Timokhin
\\[5mm]
\it Sternberg Astronomical Institute, \\
\it Universitetskij pr. 13, 119992 Moscow, Russia\\
\it e-mail: atim@sai.msu.ru
}

\date{}
\maketitle

\thispagestyle{empty}

\begin{abstract}
  \noindent Pulsar "standard model" of rotating magnetized conducting sphere
  surrounded by plasma is generalized in its essential parts for the
  case of oscillating star. Goldreich-Julian charge density,
  electromagnetic energy losses as well as polar cap scenario of
  particle accelerations are considered. Despite similarities, there
  are substantial differences between magnetospheres of rotating and
  oscillating stars. Distortion of particle acceleration mechanism in
  radiopulsars by neutron star oscillations, excited by a strong
  glitch, for example, is discussed.
 
\end{abstract}

\section{Introduction}

Neutron stars (NS) are probably the most densest objects in the
Universe. It is of the great interest for fundamental physics to study
their internal structure, because the matter inside the NS is under
extreme physical conditions (huge density, pressure, magnetic fields),
which can not be reproduced in laboratory conditions, at least in the
near future. The only way to get information about the internal
structure of a celestial object is to study its proper oscillations,
i.e doing some king of seismology. Such investigations would be in
principle possible if two things are present: an effective mechanism
for excitations of oscillations and a physical process causing changes
in radiation of the celestial object due to its oscillations. NS are
observed both as isolated objects (radiopulsars) and as members of
binary systems (X-ray binaries). In the second case it will be very
difficult to connect some observational features with proper
oscillations of the NS itself, because there are a bunch of
non-stationary processes connected with accretion of matter on to NS.
Because of this, revealing NS oscillations in radiation of
radiopulsar seems to be more promisingly. In the latter case glitches
could play a role of excitation mechanism for neutron star oscillations. Almost
the whole pulsar radiation is formed in the magnetosphere, so in order
to be able to reveal signatures of NS oscillation in the pulsar
radiation it is necessary to know how pulsar magnetosphere will be
distorted by star oscillations. In this work we are considering the
magnetosphere of an oscillating neutron star and trying to reveal
physical mechanisms able to produce observable signatures of stellar
oscillations. We consider the main properties of oscillating NS
magnetosphere and compare them with the similar properties of
magnetosphere of a rotating star.

\section{Goldreich-Julian charge density}
As it was firstly shown by Goldreich and Julian \cite{GJ},
magnetosphere of a NS is filled with charged particles. In the last
thirty years model of rotating magnetized conducting sphere
surrounded by non-neutral plasma became the standard model for
radiopulsars. Despite difficulties of this model \cite{Beskin} it is
the only model describing, at least qualitatively, main properties of
pulsar phenomena. According to this model, charge density of the
magnetospheric plasma compensates the longitudinal electric field
caused by NS rotation. The knowledge of this, so-called
Goldreich-Julian, charge density allows one to study particle
acceleration and emission mechanism. In the case of oscillations,
charge density in the magnetosphere should compensate longitudinal
electric filed caused by motion of the NS surface. So, in order to
make prediction about observational signatures of stellar oscillations
it is necessary to generalize standard pulsar model for the case of
arbitrary oscillations of the NS. Formalism of such generalization was
proposed by Timokhin, Bisnovatyi-Kogan and Spruit
\cite{Timokhin:2000}.

We have looked for configuration of electric field $\vec{\mathcal{E}}$ in
the magnetosphere of oscillating NS such that it is orthogonal to the
NS's magnetic field $\vec{B}$:
\begin{equation}
\vec{\mathcal{E}} \cdot \vec{B} \equiv 0
\label{EdotB}
\end{equation}
everywhere in the magnetosphere. Charge density supporting this
electric field is the Goldreich-Julian (GJ) charge density
connected with NS oscillations:
\begin{equation}
\rho_{\mbox{\tiny GJ}} = - \frac{1}{4\pi} \nabla \cdot \vec{\mathcal{E}}.  
\end{equation}
In \cite{Timokhin:2000} solution for GJ charge density for toroidal
oscillation modes has been constructed. It turned out that for a half
of all oscillation modes smooth charge density distribution in the
whole magnetosphere could exists only if a strong current with density
\begin{equation}
j \simeq \rho_{\mbox{\tiny GJ}} c \left(\frac{c}{\omega r}\right)
\label{current}
\end{equation}
is flowing along closed magnetic field lines ($\rho_{\mbox{\tiny GJ}}$
is the GJ charge density associated with \textit{oscillations},
$\omega$ is oscillation frequency, $r$ -- radius in spherical
coordinate system, and $c$ is the speed of light). In this work we have
constructed solution for GJ charge density for spheroidal oscillation
modes, covering the most interesting physical oscillation modes
(\textit{p-,g-,r-} modes).  Similar to the case of toroidal
oscillations, only a half of all spheroidal modes have smooth solution
without presence of a strong current flowing in the magnetosphere.
This has an important implication for possible observations of NS
oscillations, because this (non-stationary) current will flow in the
closed field lines regions, which are invisible in radiopulsars.
Hence, radiation could be produced in initially 'dead' regions of
pulsar magnetosphere. On the other hand, this will cause very rapid
damping of modes producing strong magnetospheric currents.

\begin{figure}[t]

  \centerline{\psfig{figure=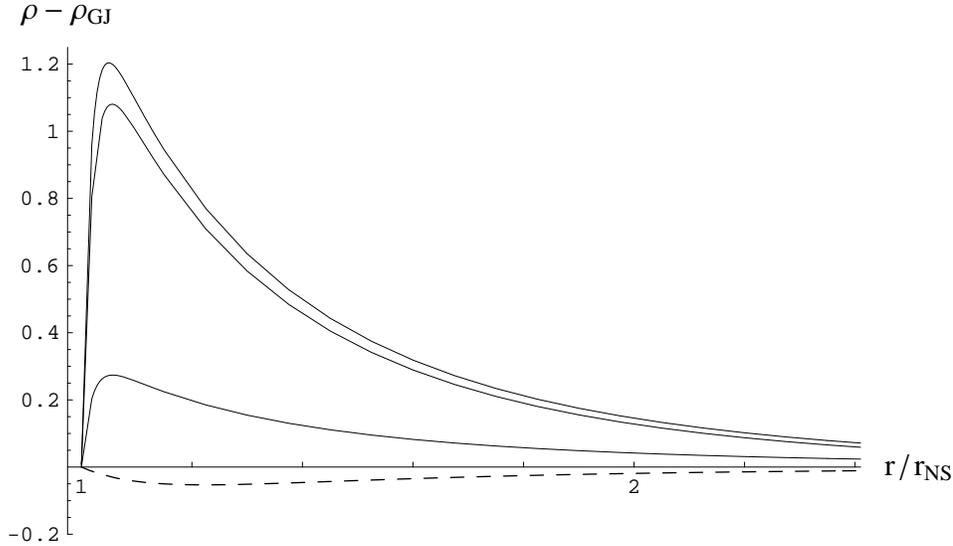,width=1.\textwidth}}
  
  \caption{Difference between the charge density of space charge
    limited flow and the local Goldreich-Julian charge density along
    magnetic field lines in the polar cap region is shown (in
    arbitrary units) for three spheroidal modes with different
    ($l,m$): (64,14) -- by the upper solid line, (54,14) -- by the
    middle solid line, (54,2) -- by the lower solid line. The same
    quantity for aligned rotator is shown by the dashed line.
    $x$-axis shows the distance $r$ from the center of the NS in units
    of the NS radius $r_{NS}$.  For each oscillation mode this
    difference is shown along the field line where a local maximum of
    corresponding $\rho_{\mbox{\tiny GJ}}$ is achieved. It was assumed
    that all modes have the same velocity amplitude and the latter is
    equal to the linear velocity at the equator of the aligned
    rotator.}
  \label{deltaRho}
\end{figure}

\section{Distortion of particle acceleration mechanism in pulsar}
In open field line regions near magnetic poles Goldreich-Julian charge
density associated with oscillations for both toroidal and spheroidal
modes with $l>1$ decreases with $r$ more rapidly than the density of
particles flowing away from the polar cap. Moreover, the difference
between charge density of outflowing particles and the GJ charge
density increases with increasing of both $l$ and $m$ (see.
Fig.~\ref{deltaRho}).  In the most popular model of space charge
limited flow, where particles freely escape the surface of the NS
\cite{Arons:Scharlemann}, this would lead to formation of a
\textit{"decelerating"} electric filed, accelerating particles back to
the NS.  If this field will be of compatible strength with
\textit{accelerating} field due to pulsar rotation, then for a half
oscillation period the resulting accelerating electric field will be
stronger or weaker than in unperturbed pulsar, when rotational and
oscillation GJ charge densities have different and the same signs
correspondingly. This will lead to variations in both spectrum and
intensity of pulsar radiation after a strong glitch. For oscillation
modes with large $l$ the corresponding GJ charge density decreases
more steeper with $r$ than GJ charge density of rotating star
\cite{Muslimov:Tsygan}. Because of this and increase of
$\rho_{\mbox{\tiny GJ}}$ with $(l,m)$, decelerating electric field
due to star oscillations with $l,m$ large enough could be of
compatible strength with the rotational accelerating field even for
surface oscillation velocity much lower than the velocity of rotation.  If
decelerating electric field is stronger than accelerating field due to
NS rotation (such a, rather artificial, case is shown in
Fig.~\ref{deltaRho}), then on field lines close to a local maxima of
$\rho_{\mbox{\tiny GJ}}$ particle acceleration will be periodically
suppressed.  To be able to make quantitative predictions about the
threshold of oscillation velocity amplitude above which oscillation of
the NS could be detected in pulsar radiation pattern one need to
perform detailed investigation of electromagnetic cascade development
in the polar cap region.  This is outside the scope of the current
work and will be subject of further investigations.

\section{Electromagnetic energy losses of oscillating star}
Oscillating star looses energy through electric current flowing
across magnetic field lines in the open filed line regions. Accepting
as an estimation for the current density 
$j\simeq \rho_{\mbox{\tiny GJ}} c$
and considering self-consistently the position of the last closed
field line we have got estimations for the electromagnetic energy
losses of an oscillating star (see \cite{Timokhin:2000}, where
detailed consideration for toroidal modes are given). In contrast to
the case of rotating star, where electromagnetic energy losses due to
current flow are well described by the magnetodipolar formula, in the case
of star oscillating in modes with $l>1$ energy losses have another
dependency on oscillation frequency than energy losses in vacuo and
depend on oscillation velocity magnitude. For rotating oscillating
star, where the position of the last closed filed line is set by
rotation, damping time of oscillations substantially differs from
estimations based on vacuum formulas too.

\section*{Acknowledgements}
This work was partially supported by
Russian Federation President Grant Program, grants
MK-895.2003.02, NSh-388.2003.2 and RFBR grant 04-02-16720

\end{document}